\documentclass[12pt, onecolumn, letterpaper]{IEEEtran}
\IEEEoverridecommandlockouts
\usepackage{amsthm}
\usepackage{amsmath,amssymb,amsfonts}

\usepackage[linesnumbered, ruled, vlined]{algorithm2e}
\usepackage{graphicx}
\usepackage{textcomp}
\usepackage{xcolor}
\usepackage{bbm}
\usepackage{cases}
\usepackage{cite}
\usepackage{subfigure}
\usepackage{tabularx}
\usepackage{bm}
\def\BibTeX{{\rm B\kern-.05em{\sc i\kern-.025em b}\kern-.08em
    T\kern-.1667em\lower.7ex\hbox{E}\kern-.125emX}}
    
\usepackage{tikz}
\newcommand*{\circled}[1]{\lower.7ex\hbox{\tikz\draw (0pt, 0pt)%
    circle (.4em) node {\makebox[0.56em][c]{\footnotesize #1}};}}
 
\newcommand{\code}[1]{{\color{black}\texttt{\small #1}}}

\begin{document}

\title{Consensusless Blockchain: A Promising High-Performance Blockchain without Consensus
}

\author{
 \IEEEauthorblockN{
      Qing~Wang, Jian~Zheng, Huawei~Huang, Jianru~Lin
    }
    
    \IEEEauthorblockN{
    Email:~ wangq79@mail.sysu.edu.cn, zhengj79@mail2.sysu.edu.cn\\
    Sun Yat-Sen University, China.
    }

 }

\maketitle

\begin{abstract}
 Consensus is unnecessary when the truth is available. In this paper, we present a new perspective of rebuilding the blockchain without consensus. When the consensus phase is eliminated from a blockchain, transactions could be canonized quickly using a well-defined universal rule without consuming hashing power. Thus, the \textit{transactions per second} (TPS) metric of such the consensusless blockchain can be largely boosted.
 Although consensus blockchain is promising, several technical challenges are also crucial. For example, double-spending attacks and frequent forking events must be prevented, the credit of block's minting must be carefully defined, and etc.
 To address those technical challenges, we propose several solutions for our consensusless blockchain (CB), including a naive monotonic scoring mechanism to calculate the ranking of each block in the chain, and a two-stage witness mechanism to add new blocks.
 The proposed CB chain is promising to offer a simplified and equipment-cheap infrastructure for rich real-world decentralized applications. 

\end{abstract}

\section{Introduction}\label{sec:intoduction} 

 %
 %
 
 Although Bitcoin's proof-of-work (PoW) \cite{nakamoto2008bitcoin} secures the transactions stored in historical blocks, it also induces extensive criticisms. 
 The most representative criticisms include its low TPS and the extraordinary amount of energy consumption spending on mining new blocks.
 %
 %
 The bitcoin's low TPS is induced by PoW consensus, in which miners compete with each other to win the opportunity of minting a new block in each round. The length expectation of each round of consensus is fixed to around 10 minutes. Even though such difficulty-defined mining cycle prevents frequent forking events, it also decides bitcoin's TPS as low as 7-10.
 %
 %
 The other shortcoming of Bitcoin's blockchain is that the PoW consensus has kept encouraging miners to upgrade their hashing capability. From CPU to ASIC-assisted mining, enterprises have developed various off-the-shelf mining hardware. For example, Nvidia's GPU cards have offered full power of crypto mining.
 The worldwide large-scale crypto mining consumes a countless amount of energy power. 
 
 %
 Besides the PoW-based consensus protocols and its variants \cite{bhosale2018volatility}, other popular consensus protocols have been proposed such as Proof of Stake (PoS) \cite{bentov2016snow}.
 Although PoS or Delegated PoS (DPoS) mechanisms are believed as environmental friendly, some argue that people who pledge large amounts of coins may have a huge influence on the consensus process, and thus affecting the decentralization of a blockchain.


 Given the criticisms aforementioned, various advanced consensus mechanisms \cite{eyal2016bitcoin, wang2019monoxide, gilad2017algorand, kiayias2017ouroboros, bentov2016snow, david2018ouroboros} have been proposed. In those consensus protocols, either transaction throughput is improved or transaction latency is reduced.
 However, those studies are basically following the direction of Proof-based consensus like Bitcoin's PoW.
 In contrast, we argue that consensus is unnecessary if we know the truth of how to resolve conflicts about all transactions and blocks. We can setup such a truth or a universal rule in a consensusless blockchain's protocol, and then the double spending problem can be tackled with a message delivery-guaranteed method instead of a complex Proof-based consensus protocol.
 That means if we design a blockchain in the consensusless way, a bunch of Proof-based consensus protocols such as PoW and its variants \cite{eyal2016bitcoin, bhosale2018volatility, wang2019monoxide}, Proof of Stake \cite{bentov2016snow, casper, david2018ouroboros}, and byzantine consensus \cite{gilad2017algorand}, can be dropped.
 %
 Previous research has demonstrated the theoretical possibility of consensusless blockchains\cite{guerraoui2019consensus}.
 
 Without consensus, TPS can be largely boosted. This is because every block node can propose new blocks frequently in a parallel way. Those new blocks are added to the chain and do not have to experience a long consensus process.

 Although Solana \cite{yakovenko2018solana} already provides a good TPS as high as 10 thousands, it still exploits consensus protocols such as proof-of-history and DPoS. Thus, the consensus of Solana is computing-intensive. 
 %
 Recently, the crypto participants are talking about the frequent outages of Solana. For example, the most recent outage of Solana was occurred on May 1st. In this outage, the Solana's blockchain was down and last for 7 hours due to the decentralized denial-of-service attacks from NFT mint bots.
 When Solana's blockchain recovered, the Solana team decided to reject any NFT minting bots. Many cryptocurrency players expressed their worries about this blasphemous censorship, which severely damages the decentralization of crypto and blockchain world.

 In contrast, our proposed CB chain uses simpler and cheaper solutions than the existing blockchains. Furthermore, our CB chain does not have a centralized censorship that degrades the spirit of decentralized world. The proposed CB chain is promising to offer a green and high-performance infrastructure to various decentralized applications (DApps) in the era of Web3.

\section{Transactions}\label{sec:tx} 

The proposed consensusless blockchain does not specify a particular transaction model. System designer can choose either the account model that is similar to Ethereum \cite{buterin2014next} or the UTXO model adopted by Bitcoin \cite{nakamoto2008bitcoin}.
In CB chain, smart contracts are supportable, and transactions could be compatible with multiple inputs and multiple outputs.
For different account models, CB chain adopts different ways to defend against double-spending attacks. CB chain has only one valid main chain. Although sometimes forks may occur, there is only one valid block at any height in the long run. Work nodes (i.e., block proposers) can verify transaction's validity by replaying all transactions storing on the entire blockchain starting from the genesis block. 
For the account-based model, double-spending attacks are defended by observing whether the \code{nonce} of two transactions are the same. For the UTXO model, double-spending attacks can be defended by verifying whether the UTXO of a transaction has been spent in another previous transaction.


\section{A Universal Rule: Ensuring the Truth of All Transactions}\label{sec:UniverseRule}

 The universe rule is set to ensure that any honest work node maintains only a single valid chain at any time.
 To achieve this goal, we need to provide two basic mechanisms, i.e., \textit{Block Scoring} and \textit{Fork Handling}.


 \subsection{Block Scoring Mechanism}
 Parameter \code{block\_score} is defined to evaluate the priority among the blocks at the same height. A block's \code{block\_score} can be calculated by a naive monotonic function using both the inputs and outputs of all the transactions packaged in the block.
 The block with lowest \code{block\_score} wins when two blocks are conflict with each other at the same height.

 \subsection{Fork Handling}
 
 Next we discuss \textit{how to address two conflict blocks} at the same height when a fork forms.
 
\begin{figure}[h!t]
    \centering 
    \includegraphics[width=0.5\textwidth]{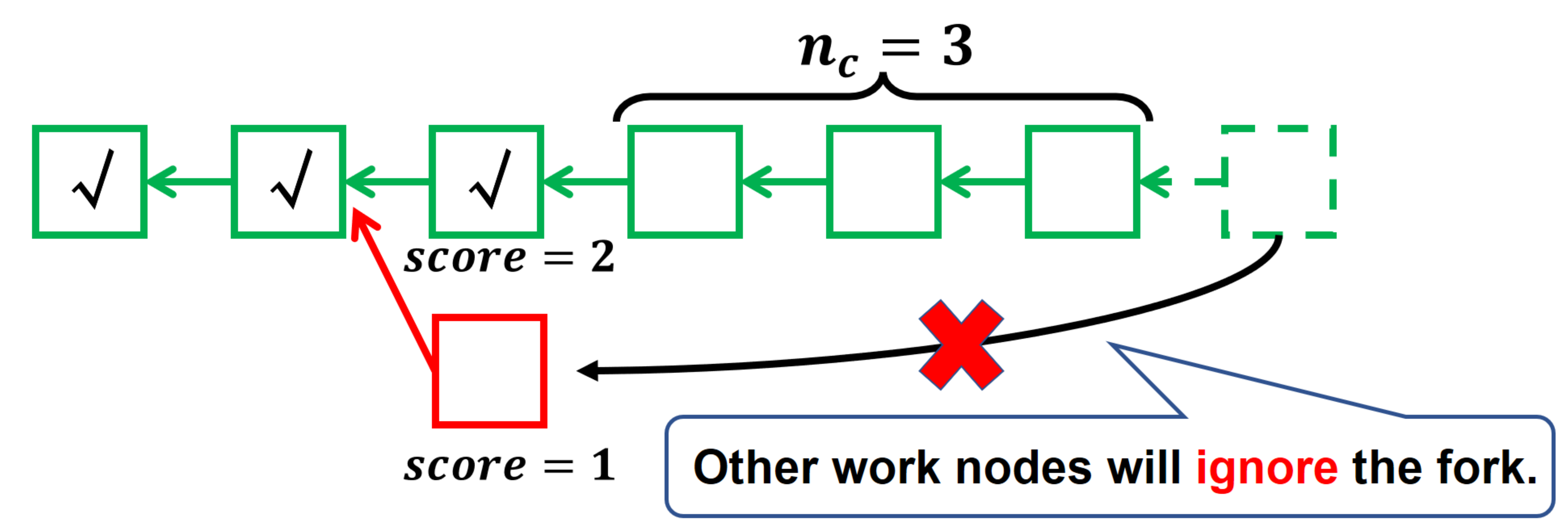}
    \caption{When $n_l \geq n_c$, work nodes need to ignore the new fork branch no matter what the \code{block\_score} of the first block is in the fork branch.}
    \label{fig:LongestChain} 
\end{figure}

\begin{figure}[h!t]
    \centering 
    \includegraphics[width=0.5\textwidth]{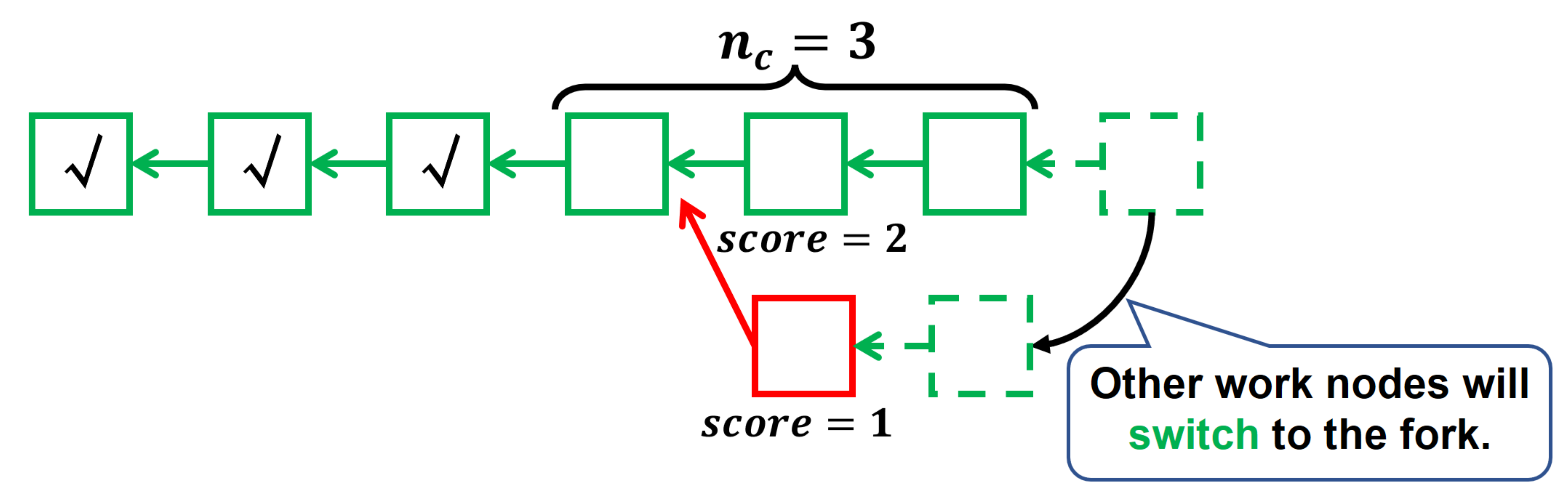}
    \caption{When $n_l < n_c$, work nodes need to compare \code{block\_score} of the first blocks in two branches.}
    \label{fig:LowestScore} 
\end{figure}

 We first define $n_c$ as the number of confirmations when a newly proposed block is admitted by the entire blockchain network.
 When a fork appears in the CB chain, we let $n_l$ denote the number of blocks in the longer fork branch. When $n_l \geq n_c$, work nodes must ignore the shorter branch as shown in Fig.\ref{fig:LongestChain}. When $n_l < n_c$, work nodes must calculate \code{block\_score} of the first blocks in two branches and choose to follow the branch whose first block has the lower \code{block\_score}, as shown in Fig.\ref{fig:LowestScore}.
 If work nodes choose to switch to the new branch of the fork, work nodes will broadcast a \code{fork-win} message to to speed up the spread of the new branch.


%

\section{Two-Stage Witness}\label{sec:2stagewitness}

When network latency is extremely low, the universal rule depicted in previous section is guaranteed. However, the inevitable network latency brings dynamics when new blocks are proposed by work nodes.
Therefore, we also propose a carefully defined \textit{two-stage witness} mechanism, aiming to enable work nodes to propose new blocks securely and fairly.
This two-stage witness includes the following 2 stages: \textit{transaction collecting} and \textit{Block Witness and Minting}. 
%


 \subsection{Collecting Transactions and Proposing a New Block} 

\begin{figure}[h!t]
    \centering 
    \includegraphics[width=0.6\textwidth]{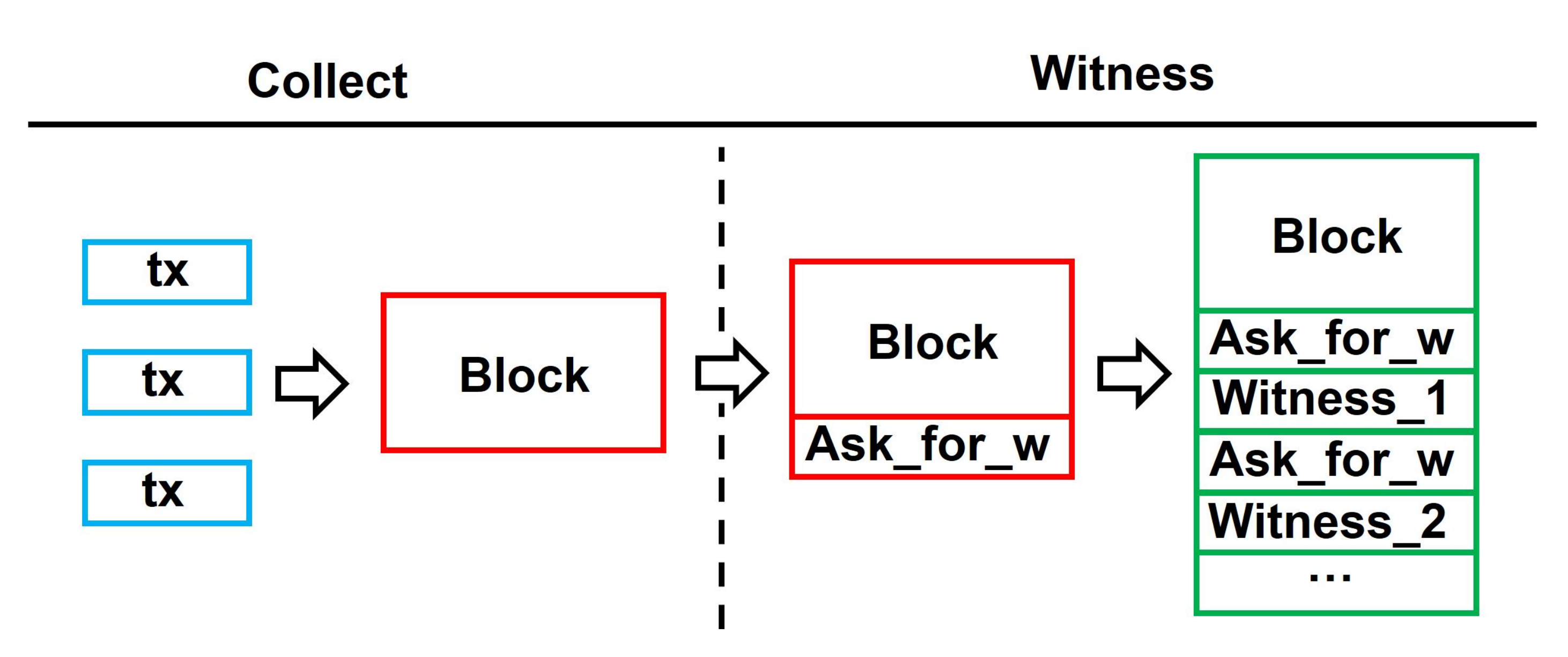}
    \caption{The proposed \textit{Two-Stage Witness}. We design a function \code{Ask\_for\_w}, aiming to allow a block's proposer to ask for endorsements (i.e., \code{witness}) from other witness nodes.}
    \label{fig:TwoStageWitness} 
\end{figure}

 As shown in Fig. \ref{fig:TwoStageWitness}, each \textit{collector} node packages a block by collecting transactions from the transaction pool. After a block is generated, every collector node can propose a new valid block following the previously-witnessed block and broadcast the new block to the blockchain network. 
 A valid block proposal needs to satisfy the following 2 conditions: 
 (i) the block includes at least a number of $\code{TX}_{\code{COUNT}}$ {\small ($\in \mathbb{N}_{+}$)} transactions, all of which are valid; and (ii) there are not other blocks at the same height having lower \code{block\_scores}.

 \subsection{Block Witness and Minting} 
 
 Work nodes continuously monitor the network. In each round of witness, each work node generates a new block by packaging valid transactions.
Then, the work node broadcasts the new block and waits for a specified number of \code{witness} signed by other witness nodes. 
%

\begin{figure}[h!t]
    \centering 
    \includegraphics[width=0.5\textwidth]{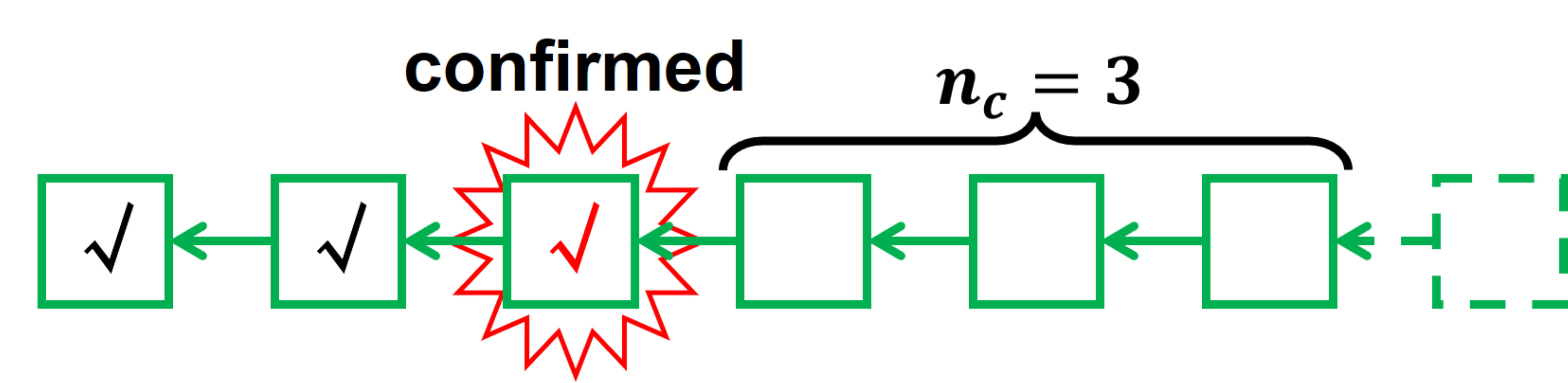}
    \caption{When a valid witnessed block receives a number of $n_c$ (\textit{e.g.,} 3) subsequent following blocks, this block is eventually confirmed.}
    \label{fig:nc} 
\end{figure}

 Every valid block must be verified by a number  $m$ of \code{witness} signatures before it is added as a valid block in the chain. 
 For any valid block, we say a \code{witness} event is successful if the following condition is satisfied. 
 \begin{align}
    \code{DistanceFunc}(\code{PK}_{\code{node}}, \code{PK}_{\code{witness}}) < \code{THRESHOLD}_{\code{witness}},
\end{align}
where the \code{DistanceFunc($\code{PK}_{\code{node}}$, $\code{PK}_{\code{witness}}$)} is a function used to select the qualified other witness nodes who can sign \code{witness} in a newly proposed block. The parameters $\code{PK}_{\code{node}}$ and  $\code{PK}_{\code{witness}}$ are the public keys of the proposer node and the witness node who provides a \code{witness}, respectively.
Parameter $\code{THRESHOLD}_{\code{witness}}$ is the customized maximum distance describing the relationship between the proposer node and another witness node who can provide a \code{witness} signature.
Thus, \code{DistanceFunc(.)} makes the block minting difficult.
When a new block receives a specified number $m$ of \code{witness} signatures from witness nodes, the block will be accepted by the entire blockchain network and other honest work nodes will follow it.
As shown in Fig.\ref{fig:nc}, when a valid witnessed block receives a number $n_c$ of subsequent following blocks, we say that this block is eventually confirmed by the CB chain.

\section{Incentive}\label{sec:incentive}

%
%
We deliberately decouple the incentive layer from CB chain to make the incentive mechanism neutral. Any particular incentive mechanism can be plugged into CB via the block mint hooks like the mechanism called \code{before\_block\_mint} and \code{after\_bolck\_mint}, which will be executed by the minter before or after a block's minting, respectively. For example, a Bitcoin-alike incentive method can be plugged in, by adding \code{coinbase} reward transactions for the collector and witnesses into the minting block in the \code{before\_block\_mint} hook. By extracting the incentive logic from the blockchain-maintenance logic, we can use CB with any economic model, even run different economic models on the same CB chain simultaneously.

We disclose more insights under the Bitcoin-alike incentive mode. Such the incentive is used to encourage collector nodes to propose new blocks. The difference is that the \code{coinbase} reward of CB chain is awarded to the proposer whose proposed new block has received the sufficient number of \code{witness} signatures.
The incentive also includes the subsidy paying to the collaborators who offer \code{witness} signatures for any newly proposed block. The incentive helps encourage work nodes to stay honest and be willing to offer \code{witness} signatures for others. If a greedy attacker is able to collude with other collaborators who can offer quick \code{witness} signatures for his new block, he would have to select his collaborators out of all work nodes. Due to the well-defined \code{DistanceFunc(.)}, it would be time-consuming to find the target collaborators. He ought to find it more profitable to play under the proposed \textit{two-stage witness} such that he can be rewarded by more new coins than the manner to collude with his collaborators.


%
%


%
%


%
%


%
%
\section{Calculations}\label{sec:calculation}

\subsection{Security Analysis of Witnessing an Adversarial Block}

To avoid Sybil attacks, every work node needs to pledge a certain amount of assets before participating the CB network.
Let $q$ {\small ($\in [0, 1]$)} represent the proportion of adversarial nodes. Considering that the selection of witness nodes is random, the probability that a qualified witness is in fact an adversarial node is also $q$. That is,
 \begin{align}
    Pr(\textit{a qualified witness node is an adversarial node}) = q.
\end{align}

We then consider an extreme adversarial case, in which an invalid block is witnessed by all adversarial collaborator nodes.
The probability of such extreme case is $q^{m}$, i.e.,
 \begin{align}\label{eq:invalidPr}
    Pr(\textit{an invalid block is witnessed}) = q^{m}.
\end{align}

Eq. \eqref{eq:invalidPr} indicates that an invalid block is almost impossible to win a successful witness when $m$ is large enough.


\subsection{The Safety of CB Chain versus Message-Delivery Ratio}

Hard forks are always undesired in a blockchain. In this part, we analyze the threat to system safety when a permanent hard fork forms. This is because a part of honest work nodes will be misled by the hard fork, when the message-delivery ratio is not high. Once any honest work node is misled by the hard fork, the safety of the blockchain will be weakened.



We then calculate the probability of the situation (denoted by $Pr_{\code{misled}}$) when an honest node is misled by a hard fork.
The related parameters include the message-delivery ratio $r$ {\small ($\in (0, 1))$}, the number of confirmation blocks $n_c$, the \code{witness} configuration parameter $m$, and the number of \code{fork-win} message $l$.
Thus, the probability $Pr_{\code{misled}}$ is written as follows.
%
 \begin{align}\label{eq:misled}
    Pr_{\code{misled}} = (1\ -\ r)^{(m+1)*(n_c+1)*(m+n_c+2)+l}.
 \end{align}


\textbf{Insight of Eq. \eqref{eq:misled}}: In a perfectly-connected network, the message-delivery ratio can approximate 100\%. Even if $r$ is not very high in a real-world blockchain network, Eq. \eqref{eq:misled} implicates that $Pr_{\code{misled}}$ can be maintained in a very low level by choosing appropriate values of parameters $n_c$ and $m$.
Considering the most conservative design(the number of confirmation blocks $n_c$ is $2$ and the \code{witness} configuration parameter $m$ is $2$), the probability of hard bifurcation of a block from generation to confirmation is less than $1/10^{54}$. Table.\ref{tab:hard_fork} shows that hard forks are almost impossible to occur naturally.

\begin{table}[!ht] 
\centering
\caption{Probability of a hard fork under the same scale of existing blockchains}
\begin{tabular}{|c|c|c|c|c|c|} \hline 
Blockchain & Height & Time of existence & Probability of a hard fork at the same scale & Expected time for a hard fork \\ \hline
Bitcoin & 751789 & 14 years	& $<1/10^{47}$ & $>10^{47}$ years   \\ \hline
Ethereum & 15437870 & 7	years & $<1/10^{45}$ & $>10^{37}$ years \\ \hline
Solana & 148287091 & 4 years & $<1/10^{44}$ & $>10^{35}$ years \\ \hline
\end{tabular}
\label{tab:hard_fork} 
\end{table}


\begin{figure}[h!t]
    \centering 
    \includegraphics[width=0.5\textwidth]{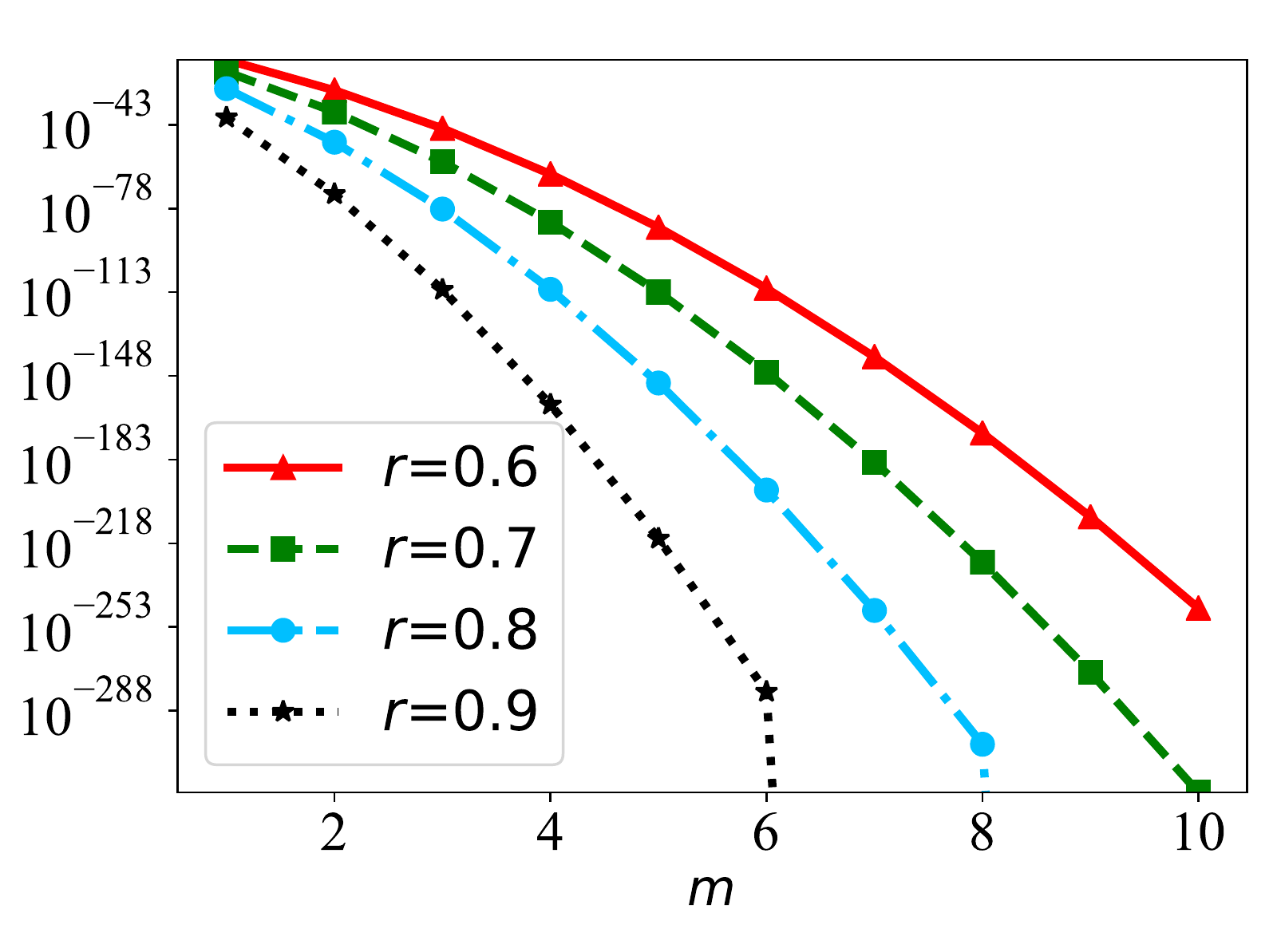}
    \caption{The probability $Pr_{\code{misled}}$ versus parameter $m$, given different message-delivery ratio $r$.}
    \label{fig:P_m} 
\end{figure}

We conduct a simulation to verify the implication of Eq. \eqref{eq:misled}. 
When $n_c$ is set to 3 and the exist of \code{fork-win} message is ignored, Fig.\ref{fig:P_m} shows that the probability $Pr_{\code{misled}}$ declines rapidly as the number of \code{witness} configuration parameter $m$ increases, under different settings of message-delivery ratio $r \in \{0.6, 0.7, 0.8, 0.9\}$. 
When $r$ is not high, CB can still obtain a very low $Pr_{\code{misled}}$ by setting a suitable number of confirmation parameter $n_c$ and choosing a suitable parameter $m$.



\section{Conclusion}\label{sec:conclusion} 

Consensusless blockchain is a game-changing design that is going to lead a new generation of blockchain protocols in the era of web3. Our proposed CB chain is promising to offer a high-performance trusted infrastructure for both research scenarios and rich real-world decentralized applications.
 
\bibliographystyle{IEEEtran}
\bibliography{reference}

\end{document}